**Preprint Version**

**Regression and Forecasting of U.S. Stock Returns Based on LSTM**

Shicheng Zhou
University of Minnesota, Minneapolis, MN 55455, United States, shicgz@gmail.com

Zizhou Zhang *
UIUC, Champaign, IL 61801, United States, zhangzizhou_2000@outlook.com

Rong Zhang
University of California, Davis, CA 95616, United States, rzhang1118@gmail.com

Yuchen Yin
Columbia University, New York, NY 10027, United States, yy3243@tc.columbia.edu

Chia Hong Chang
University of Colorado Boulder, CO 80309, United States, chch2712@colorado.edu

Qinyan Shen
Independent researcher, Jersey City, NJ07302, United States, qinyan.shen28@gmail.com

**Abstract**

This paper analyses the investment returns of three stock sectors, Manuf, Hitec, and Other, in the U.S. stock market, based on the Fama-French three-factor model, the Carhart four-factor model, and the Fama-French five-factor model, in order to test the validity of the Fama-French three-factor model, the Carhart four-factor model, and the Fama-French five-factor model for the three sectors of the market. French five-factor model for the three sectors of the market. Also, the LSTM model is used to explore the additional factors affecting stock returns. The empirical results show that the Fama-French five-factor model has better validity for the three segments of the market under study, and the LSTM model has the ability to capture the factors affecting the returns of certain industries, and can better regress and predict the stock returns of the relevant industries.

## 1 Introduction

In recent years, with the rapid development and complexity of financial markets, the traditional capital asset pricing model (CAPM) is no longer able to fully capture the risk factors in the market and predict stock returns. Since Fama and French proposed the three-factor model in 1992, scholars have been striving to extend and refine these models to improve the accuracy and explanatory power of stock return forecasts. The three-factor model was subsequently extended by Carhart into a four-factor model by adding momentum factors, and eventually evolved into the five-factor model proposed by Fama and French in 2015 [1].

Although these models have gained wide application and recognition in academia and practice, their effectiveness in dealing with data from some specific sectors or emerging markets remains to be verified. With the development of artificial intelligence technology, machine learning methods, especially those based on neural networks, show potential in the field of financial forecasting[2-3]. Ke et al. (2024) utilized BP-GA to study the volatility and returns of indices, achieving commendable results. This has served as a valuable inspiration for us to explore neural networks after employing traditional models [4]. Meanwhile, Hu et al. (2024) successfully applied GANs models in their research on Bitcoin returns, which not only confirmed the effectiveness of using the latest models but also pointed the way for our future research endeavors[5]. LSTM models, as a special kind of recurrent neural network, have attracted much attention for their effectiveness in dealing with long-term dependence problems in time series data， like prediction of return and risk management[6-10]. Of course, GANs is also another popular approach[11].

The purpose of this study is to compare the predictive effectiveness of three-factor, four-factor, and five-factor models in three major U.S. stock sectors through empirical analyses, and to explore the application of LSTM models in stock return prediction in order to provide insights into the integration of traditional financial theories with modern machine learning methods. This study aims to answer the following core questions: Comparison of model effectiveness: How effective are the three-factor, four-factor, and five-factor models in predicting stock returns for each of the major U.S. stock market sectors? Are there any significant differences in the performance of these models across different industry sectors? Application of LSTM models: Can LSTM models provide predictive accuracy beyond that of traditional factor models in stock return forecasting? What are the strengths and possible challenges of LSTM models when dealing with stock market data? Possibilities for model fusion: Is it possible to combine LSTM models with traditional factor models to predict stock returns? Possibilities of model fusion: Is it possible to combine traditional factor models with LSTM models to improve the accuracy of stock return forecasting?

In order to answer the above research questions, the following analytical steps were taken in this study:

Data collection and processing: Data source: The databases of the three major U.S. stock markets (NYSE, AMEX, and NASDAQ) are selected as the main data source to collect monthly data from January 2004 to January 2024, including stock returns and related financial variables of the Manuf, Hitec, and Other sectors[12-13]. Pre-processing: the collected data are cleaned, including the filling of vacant values and the elimination of outliers, to ensure that the quality of the data meets the requirements of statistical analyses and machine learning models.

Model building and validation： Construct the three-factor, four-factor and five-factor models of Fama-French, and calculate the corresponding regression coefficients and statistical significance[14]. LSTM model building: design the LSTM network architecture, determine the appropriate hyperparameters, and divide the dataset into the training set and the test set, which are used to train and validate the predictive performance of the model.

Model comparison： The goodness of fit and prediction accuracy of the models were assessed by comparing the statistical metrics such as R-squared, RMSE (root mean square error) and MAE (mean absolute error) of the different models.

Analysis and interpretation of results: To analyse the contribution of factors in different models and their explanatory power for industry stock returns. Model Effectiveness: Discuss the performance of LSTM models in predicting stock returns and their advantages and limitations relative to traditional factor models[15-16]. Industry-specific analysis: to explore the differences in the performance of the models in different industry sectors and analyse how industry characteristics affect the predictive effectiveness of the models.

Conclusions and outlook： Summary: To summarise the research findings and summarise the strengths and limitations of each model. Directions for future research: based on the results of this study, suggest possible directions for future research, such as exploring the application of other machine learning techniques in stock market forecasting.

## 2 RESEARCH METHODOLOGY AND DATA

### 2.1 Three-factor, four-factor, and five-factor models

Fama and French (1992) showed empirically that size and book-to-market ratio significantly affect stock returns, and then proposed a three-factor model based on the CAPM model[17-18]; on the basis of the three-factor model of Fama and French (1993), CARHART (1995) added one-year return momentum anomalies and constructed a four-factor model[19-20]. However, scholars in subsequent empirical studies found that the three risk factors could not fully explain all the excess returns[21-25], and Fama and French (2013) constructed a five-factor model based on the three-factor model with the addition of profitability factor and investment factor. Variables definition are as below Table 1.

**Table 1 Interpretation of variables**

| Parameters of an equation | Explanation |
|---|---|
| Rmkt-Rf(in F-F3, Car4, F-F5 modles) | Excess returns in the market |
| SMB(in F-F3, Car4 modles) | Average return on three small portfolios minus average return on three large portfolios |
| SMB(in F-F5 modles) | Average return of six small portfolios minus average return of six large portfolios |
| MOM(in Car4 modles) | Average return of two high-prior-return portfolios minus average return of two low-prior-return portfolios |
| HML(in F-F3, Car4, F-F5 modles) | Average return of two value portfolios minus average return of two growth portfolios |
| RMW(in F-F5 modles) | Average return of two solidly profitable portfolios minus the average return of two portfolios with weaker operating profitability |
| CMA(in F-F5 modles) | Average return of two conservative portfolios minus average return of two aggressive portfolios |

### 2.2 Multiple Linear Regression Models

A multiple linear regression model describes how the dependent variable Y varies with changes in multiple independent variables X. The model is a linear regression model. Assuming that the value of the dependent variable Y relative to the m independent variables X1, X2, ...... Xm are determined for each of the n observations, the general form of the multiple linear regression model is:

$$Y = \beta_0 + \beta_1 X_1 + \beta_2 X_2 + \ldots + \beta_n X_n + e \qquad (1)$$

$\beta_0$ in equation (1) is the constant term, also known as the intercept. $\beta_1$, $\beta_2$, ...... , $\beta_m$ are called partial regression coefficients, or simply regression coefficients. The equation shows that the response variable Y in the data can be approximated as a linear function of the independent variables $X_1$, $X_2$, ...... , and $X_m$ as linear functions. The partial regression coefficient $\beta_j$ (j=1, 2, ... , m) represents the average amount of change in Y when X is changed by one unit while the other independent variables are held constant.

### 2.3 LSTM

Recurrent Neural Networks (RNN) are widely used in the fields of epidemiological transmission, environmental monitoring, financial markets, etc[26-27]. The Long Short-Term Memory Neural Network (LSTM), which is improved based on the RNN algorithm, solves the problems of gradient disappearance, gradient explosion and long term dependence that occur in RNN in analysing and predicting the time series data and identifying the changing patterns of the time series data[28-31]. The LSTM, through the introduction of the memory unit and the gating mechanism, effectively captures complex patterns in sequences and shows good performance on non-smooth sequences[32-38]. The LSTM model memory unit consists of three parts, namely the forgetting gate, input gate and output gate.

### 2.4 Data

The data processing of the experimental data of this study mainly includes the filling of vacancy values, the elimination of outliers and data statistics. For the vacant values, the Lagrange interpolation method is used for processing, and for the outliers, the outliers are eliminated and processed in a similar way to the vacant values.

## 3 Results

### 3.1 Multiple Linear Regression Model

#### 3.1.1 *Manuf.*

Within the Manuf plate, the data of interest were regressed using a multiple linear regression model. F-F3 regression results:

$$R_i - R_f = 0.9219(R_{mkt} - R_f) + 0.0473SMB + 0.033HML \qquad (2)$$

**Table 2** **Factor significance in F-F3 in Manuf**

| Divisor Coefficient | P-value |
| --- | --- |
| $R_{mkt} - R_f$ | <2e-16*** |
| SMB | <2e-16*** |
| HML | 1.11e-15*** |

*** indicates p-value < 0.001, highly significant; ** indicates p-value < 0.01, significant; * indicates p-value < 0.05, statistically significant.

Carhart4 regression results:

$$R_i - R_f = 0.905(R_{mkt} - R_f) + 0.05SMB + 0.009HML - 0.06MOM \quad (3)$$

**Table 3** **Factor significance in Carhart4 in Manuf**

| Divisor Coefficient | P-value |
| --- | --- |
| $R_{mkt} - R_f$ | <2e-16*** |
| SMB | <2e-16*** |
| HML | 1.70e-11*** |
| MOM | 3.93e-10*** |

*** indicates p-value < 0.001, highly significant; ** indicates p-value < 0.01, significant; * indicates p-value < 0.05, statistically significant.

F5 regression results:

$$R_i - R_f = 0.95(R_{mkt} - R_f) + 0.08MB + 0.02HML + 0.1RMW + 0.01CMA \quad (4)$$

**Table 4** **Factor significance in F-F5 in Manuf**

| Divisor Coefficient | P-value |
| --- | --- |
| $R_{mkt} - R_f$ | <2e-16*** |
| SMB | <2e-16*** |
| HML | 0.000596*** |
| RMW | 0.420 |
| CMA | 0.859 |

*** indicates p-value < 0.001, highly significant; ** indicates p-value < 0.01, significant; * indicates p-value < 0.05, statistically significant.

The regression superiority of each model was evaluated:

**Table 5** **Regression Superiority in Manuf**

| Manuf | R-squared | P-value | RMSE | MAE |
|---|---|---|---|---|
| F-F3 | 0.901 | 1.12e-94*** | 1.514 | 1.161 |
| Carhart4 | 0.904 | 1.52e-94*** | 1.490 | 1.159 |
| F-F5 | 0.909 | 2.59e-95*** | 1.452 | 1.120 |

*** indicates p-value < 0.001, highly significant; ** indicates p-value < 0.01, significant; * indicates p-value < 0.05, statistically significant.

As can be seen from the data in Tables 2, 3, 4 and 5, the coefficients of all three factors in the F-F3 model are highly significant, with p-values much less than 0.001. In the Carhart4 model, the coefficients of the four factors are also highly significant, with p-values much less than 0.001. In the F-F5 model, the coefficients of the first three factors are significant (with a p-value of 0.000596 for the third factor, which is still very small), but the coefficients of the last two factors are not significant, with p-values of 0.420 and 0.859 respectively, which means that their effects on the dependent variable are significant. factor has a p-value of 0.000596, which is still very small), but the coefficients of the last two factors are not significant, with p-values of 0.420 and 0.859 respectively, which implies that their effects on the dependent variable are not statistically significant; at the same time, regression superiority of all three models performs well. In summary, all models show high regression superiority and coefficient significance, with the F-F5 model slightly outperforming the other two models in terms of explaining variation and predictive accuracy. Thus, in this paper, the multiple linear regression equation derived using the five factor model will be used as the estimating equation for the rate of return in this industry:

$$R_i = R_f + 0.95(R_{mkt} - R_f) + 0.08MB + 0.02HML + 0.1RMW + 0.01CMA \quad (5)$$

#### 3.1.2 *Hitec.*

The regression superiority of each model was evaluated:

**Table 6** **Regression Superiority inHitec**

| Hitec | R-squared | P-value | RMSE | MAE |
|---|---|---|---|---|
| F-F3 | 0.864 | 1.37e-81*** | 1.831 | 1.512 |
| Carhart4 | 0.864 | 2.97e-80*** | 1.831 | 1.512 |
| F-F5 | 0.871 | 2.84e-81*** | 1.780 | 1.462 |

*** indicates p-value < 0.001, highly significant; ** indicates p-value < 0.01, significant; * indicates p-value < 0.05, statistically significant.

As can be seen from the data in Tables 6, the regression superiority of all three models performs well. Taken together, despite the high regression goodness of all models, the F-F5 model performs slightly better in terms of regression goodness, especially in terms of prediction accuracy. However, it is worth noting that one of the factors in the F-F5 model has an insignificant coefficient, which may mean that it is not important in explaining the variation in the dependent variable. Thus, this paper will use the multiple linear regression equation derived from the five-factor model as an estimator for the return of this industry:

$$R_i - R_f = 0.9(R_{mkt} - R_f) + 0.02MB - 0.19HML - 0.08RMW - 0.06CMA \quad (6)$$

#### 3.1.3 *Other.*

The regression superiority of each model was evaluated:

**Table 7** **Regression Superiority in Other**

| Other | R-squared | P-value | RMSE | MAE |
|---|---|---|---|---|
| F-F3 | 0.936 | 3.10e-122*** | 1.283 | 0.957 |
| Carhart4 | 0.940 | 1.73e-113*** | 1.240 | 0.925 |
| F-F5 | 0.946 | 3.38e-116*** | 1.178 | 0.896 |

*** indicates p-value < 0.001, highly significant; ** indicates p-value < 0.01, significant; * indicates p-value < 0.05, statistically significant.

As can be seen from the data in Tables 7, the p-values for all factors in the F-F3 model are much less than 0.001, indicating that all factor coefficients are highly statistically significant. the p-values for all factors in the Carhart4 model are much less than 0.001, again indicating that all factor coefficients are highly statistically significant. the F-F5 model has p-values for all factors that are much less than 0.001, indicating that all factor coefficients in this model are highly significant;

Meanwhile, the regression superiority of all three models performs well. Overall, all models are very strong in terms of statistical significance of factor coefficients, and there is no problem of lack of significance. In terms of explanatory power and predictive accuracy, the F-F5 model performs the best, followed by the Carhart4 model, while the F-F3 model is relatively weak. Thus, in this paper, the multiple linear regression equation derived using the five-factor model will be used as the estimating equation for the rate of return in this industry:

$$R_i - R_f = 0.83(R_{mkt} - R_f) - 0.03MB + 0.2977HML - 0.1RMW - 0.07CMA \quad (7)$$

In summary, an investor can substitute the values of, SMB, HML, RMW, CMA predicted for a particular month in the future to calculate the industry's predicted industry yield for that month.

### 3.2 LSTM

Using LSTM for regression on relevant data, with the training set and test set split in a 7:3 ratio.

**Table 8 LSTM regression superiority**

| | R-squared | RMSE | MAE |
|---|---|---|---|
| Manuf | 0.903 | 1.470 | 1.121 |
| Hitec | 0.929 | 1.888 | 1.525 |
| Other | 0.909 | 1.531 | 1.149 |

Comparing the corresponding data in Table 8, because the Fama-French five-factor model is an extension of the Capital Asset Pricing Theory, when this model shows a high R-squared, it indicates that the model has been able to effectively capture the systematic risk factors that affect stock returns.

In the cases of the Manufacturing and Other sectors, if the five-factor model already has a coefficient of determination over 0.9, this may imply that the selected factors have largely explained the variations in these industries' returns[39-40]. Such a model already possesses excellent explanatory power, and further use of LSTM may not significantly enhance predictive performance. Moreover, from the perspectives of computational efficiency and model interpretability, a simpler model may be a better choice in this scenario.

For the Hitec sector, the relatively low coefficient of determination of the five-factor model indicates some limitations in capturing the variations in industry returns. This could be due to industry-specific risk factors or influences from market microstructure that are not adequately captured by the five-factor model. In this context, introducing an LSTM model may be more valuable[41-43]. Since LSTM networks can handle long-term dependencies in time series data and can learn more complex nonlinear patterns from the data, they may provide an effective way to abstract predictive signals from historical information, thereby enhancing the ability to capture industry-specific influencing factors[44].
In practice, for financial time series data with highly nonlinear characteristics and complex market dynamics, machine learning methods like LSTM may offer an effective alternative, especially when traditional linear models fail to capture all relevant information adequately[45]. However, the application of machine learning models also brings about reduced model interpretability and risks of overfitting.

## 4 Conclusion

This study conducts predictive analysis on the stock returns of the Manufacturing (Manuf), High Technology (Hitec), and Other sectors in the US stock market by comparing the Fama-French three-factor, four-factor, and five-factor models, combined with an LSTM model. The main conclusions are as follows:

Model Applicability: The Fama-French five-factor model generally demonstrates high predictive accuracy and explanatory power across the three stock sectors, particularly in the Manufacturing and Other sectors, where their R-squared values exceed 0.9. This suggests that the model effectively captures the systematic risk factors influencing stock returns.

Factor Importance: Cross-sector comparisons indicate that Rmkt-Rf and SMB consistently show significance across all models, while RMW and CMA factors exhibit insufficient significance in certain industry models. This may imply variations in the importance of these factors across different industries.

Predictive Performance: In the Hitec sector, the LSTM model demonstrates superior predictive ability compared to traditional factor models. This suggests that the LSTM model has potential advantages in handling data with complex market dynamics and highly nonlinear characteristics.

Potential for Model Fusion: Despite the impressive performance of the LSTM model in certain scenarios, its high data requirements and computational complexity might make it more suitable for practical applications when combined with traditional factor models.

Investment Strategy: Investors should consider using the five-factor model as the primary tool for evaluating stocks in the Manufacturing and Other sectors. However, in the High Technology sector, combining the LSTM model may provide additional predictive advantages.

Future research could consider introducing a wider variety of machine learning algorithms, such as Convolutional Neural Networks (CNN) and attention mechanisms, to explore their applications in stock market prediction. Additionally, conducting similar analyses on data from more industries or emerging markets may reveal new insights into model performance under different market structures.